%Paper: astro-ph/9403045
%From: nemiroff@grovx1.DNET.NASA.GOV
%Date: Tue, 22 Mar 94 17:47:48 -0500

% This is Plain TeX, Version 3.0
% 14 pages, 17 figures (available by FAX)
% *** Macro Free Zone ***

\magnification=1200
\baselineskip=14pt

\def\cl{\centerline}

\ \

\bigskip\bigskip
\bigskip\bigskip
\bigskip\bigskip

\cl{\bf NULL RESULT IN GAMMA-RAY BURST LENSED ECHO SEARCH }

\bigskip\bigskip
\bigskip\bigskip

\cl{ R.J. Nemiroff,$^{(1),(2),(3)}$ W.A.D.T. Wickramasinghe,$^{(4)}$ }
\cl{ J.P. Norris,$^{(3)}$ C. Kouveliotou,$^{(1),(5)}$ G.J. Fishman,$^{(5)}$ }
\cl{ C.A. Meegan,$^{(5)}$ W.S. Paciesas,$^{(5),(6)}$ and J. Horack$^{(5)}$ }

\bigskip\bigskip
\bigskip\bigskip

\cl{\it $^{(1)}$ Universities Space Research Association }
\cl{\it $^{(2)}$ George Mason University, CSI, Fairfax, VA  22030 }
\cl{\it $^{(3)}$ NASA Goddard Space Flight Center, Greenbelt, MD 20771 }
\cl{\it $^{(4)}$ Dept. Astron. \& Ap., University of Pennsylvania, PA 19104 }
\cl{\it $^{(5)}$ NASA Marshall Space Flight Center, Huntsville, AL 35812 }
\cl{\it $^{(6)}$ University of Alabama, Huntsville, AL 35899 }

\bigskip\bigskip
\bigskip\bigskip
\bigskip\bigskip
\bigskip\bigskip
\bigskip\bigskip
\bigskip\bigskip

\cl{ In press: The Astrophysical Journal }
\cl{ Scheduled for the 10 September 1994 issue }

\vfill\eject
\baselineskip=18pt

\ \
\bigskip\bigskip

\cl{\bf ABSTRACT }

\bigskip

We have searched for gravitational-lens induced echoes between gamma-ray
bursts (GRBs) in BATSE data. The search was conducted in two phases.  In
the first phase we compared all GRBs in a brightness complete sample of the
first 260 GRBs with recorded angular positions having at least 5 \% chance
of being coincident from their combined positional error.  In the second
phase, we compared all GRB light curves of the first 611 GRBs with recorded
angular positions having at least 55 \% chance of being coincident from
their combined positional error. No unambiguous gravitational lens
candidate pairs were found in either phase, although a ``library of close
calls" was accumulated for future reference.  This result neither excludes
nor significantly constrains a cosmological origin for GRBs.

\noindent {\it Subject headings:} cosmology, gamma-rays: bursts,
gravitational lensing

\vfill\eject

\cl{\bf 1. INTRODUCTION }

\bigskip

There is no present consensus as to the cause of or distances to gamma-ray
bursts (GRBs). Recent results by the Burst and Transient Source Experiment
(BATSE) on board the {\it Compton} Gamma Ray Observatory have confirmed
their isotropic angular distribution and ``confined" peak brightness
distribution (Meegan et al. 1992). These results generate an interesting
puzzle.  Currently, there are more than 100 models for the origins of GRBs
(Nemiroff 1993).  To help eliminate inapplicable models, discriminating
statistical tests are needed.

The idea of searching for a gravitational lens echo in gamma-ray burst data
is not a new one.  Paczynski (1986b, 1987) originally suggested that such a
lens echo effect might be detectable were GRBs to lie at cosmological
distances.  Mao (1992) estimated that the number of BATSE GRBs that need to
be inspected to find a gravitational lens echo would be between 250 and
2000. Grossman and Nowak (1994) estimate that BATSE should detect a lensing
event every 1.5 - 25 years, depending on the cosmological scenario. Blaes
and Webster (1992) pointed out that GRB echoes would be expected if the
universe were populated with a significant fraction of compact objects near
$10^6$ solar masses. Narayan and Wallington (1992) discussed what
information about the lens might be derived from the discovery of a
gravitational lens echo.

A preliminary echo search of the first 386 GRBs detected by BATSE was
conducted by Nemiroff et al. (1993a) without success. A study searching for
lens effects within the time stream of the first 44 BATSE GRBs for massive
compact dark matter (Nemiroff et al. 1993b) also did not find an echo but
was able to show that either GRBs do not occur at the most likely
cosmological distance implied by their brightness distribution or that a
universe composed to closure density of compact objects with masses
between $10^{6.5}$ and $10^{8.1}$ solar masses is marginally excluded.

A gravitational lens echo is to be expected if a galaxy is near enough to
the line of sight to a GRB to create more than one detectable image of the
GRB.  For galaxies such images would be separated by at most several
arcseconds, just as lensed QSO images are.  However, the time delay between
images can be from hours to years, depending on the lens geometry and the
relative placement of the source behind the lens.  For a canonical
isothermal galaxy lens involved in QSO lensing, the time delay between the
two brightest images is on the order of a few  months, with one or two
years possible when a cluster of galaxies is superposed in the field
(Blandford and Narayan 1992).  Galaxies of smaller mass, which are more
numerous, are capable of creating sub-arcsecond images, and time delays on
the order of a week might be expected (Grossman and Nowak 1994). For lens
geometries where more than one image is expected, the shortest time between
images would be on the order of several hours, while again one or two years
is roughly the maximum amount of time expected between images (Narasimha
1993). In general the brighter an image pair, the shorter the time delay
between images, and the less probable an observer-galaxy-source alignment
would be that would create this bright pair.  Images themselves created by
a cluster lens would typically be expected to be separated by 10 years or
more - a time period too long to search for in current BATSE data.

In general, one cannot hope to resolve GRB images angularly - one can only
hope to resolve them temporally.  Both the random and systematic errors in
the positions to BATSE GRBs, which are on the order of degrees (Brock et
al. 1992), are much greater than the expected angular separation of images
created by a galactic lens effect.  Therefore only GRB events with
positions that could be coincident will be searched as
candidates for being gravitational lens images of the same burst.

Gravitational lens image pairs would be expected to have time profiles that
are identical to within a scale factor.  This assumes that any beaming
inherent in the GRB progenitor scenario cannot be collimated to much better
than a few arcseconds, as this is a likely angular size for the galaxy lens
(at least the part that would be involved in a lens effect) on the GRB's sky.

By the equivalence principle, gravitational fields bend the light of
different energies equally.  Therefore, gravitational lens images would also
be expected to have identical spectra. More stringently, since it is known
that the spectra of most GRBs change over the duration of the burst
(Norris et al. 1985), lensed images would be expected to have identical
spectra at all times during the burst.

As gamma rays traverse most material without absorption, there is not
expected to be any significant absorption between images.  One can not,
then, rely on such a mechanism to alter one images' spectra relative to the
other.  It is even harder to imagine a scenario where the amount of
absorption changes over the time-scale of the GRB, which would be needed to
alter the light curve.

Our present search differs from the lens search and subsequent dark matter
limits discussed in Nemiroff et al. (1993b) in that here we are comparing
different GRB light curves to each other in a search for galactic lenses,
while Nemiroff et al. (1993b) searched the time stream immediately
following GRBs for an echo indicative of lower mass dark matter lenses. In
\S 2 we describe the search for gravitational lens echoes and the
statistical method used to distinguish an echo. In \S 3 we present the
results of the search and we discuss these results and inferences in \S 4.

\bigskip

\cl{\bf 2. THE SEARCH FOR GRAVITATIONAL LENSING }

\bigskip

Our search for lensed echoes occurred in two phases.  In Phase I only the
first 260 GRBs were considered, as these GRBs had calibrated peak fluxes. Of
these 260 GRBs we used the same peak-flux complete sample of 118 GRBs as
Wickramasinghe et al. (1993). This sample is composed of GRBs above a
limiting peak flux level to which the sample is 99\% complete, and demands
that the GRBs were measured during a time of normal background.

The angular positions of these 118 were compared, and those pairs with
recorded angular positions having at least 5 \% chance of being coincident
considering their combined positional error were retained for further
analysis. This left 104 pairs of GRBs for the Phase I comparison.

For the Phase II search, the first 611 GRBs were considered. This
incorporated all GRBs detected by BATSE before and including 6 April 1993,
with trigger numbers from 105 to 2291. The positions of all 611 were
compared and those pairs that had recorded angular positions having at
least 55 \% chance of being coincident considering their combined
positional error were kept.  We excluded those GRB pairs with angular
separations greater than 30$^o$, which in general were extremely weak. This
weakness created a large angular uncertainty in the GRB's position which in
turn created an inordinately large number of potentially coincident GRB
image pairs.  In addition, many of these GRBs were so weak that they would
never be very convincing in a statistical comparison.  This left 1706 pairs
of GRBs for the Phase II comparison.

The initial comparison procedure was visual.  A hard copy of the light
curve in channel 3 (100-330 keV) of each GRB was made and catalogued.
Channel 3 was chosen because it usually has the highest signal and
relatively low and featureless background. A single energy channel was
chosen for the initial comparison so that other energy channels could
potentially be used later for an independent comparison.

Light curves that bore a marked similarity were recorded for future
numerical comparison. For a wide range of scale factors and time offsets,
those pairs of recorded GRBs were compared via a modified $\chi^2$ test to
see if they could both have been drawn from the same parent distribution. A
similar statistical comparison test has been suggested by Wambsganss
(1993).  A statistic based on the Fourier transform is given in Nowak and
Grossman (1994).

The $\chi^2$ statistic was computed as follows.  A background level for
each GRB was computed, and a subjective starting and stopping time was
determined. The stronger GRB (designated \#1 here) was shifted in time
$\Delta t$ (usually in units of the 64 ms time bins) and decreased in flux
by an amount $f$, in each bin.  The dimmer GRB was designated \# 2. For a
wide range of $f$ and $\Delta t$ values, the running sum $\chi^2_{12} =
\Sigma (f*(C_1-B_1) - (C_2-B_2))^2$ was computed, as well as running sums of
$\chi^2_{1B} = \Sigma (C_1 - B_1)^2$ and $\chi^2_{2B} = \Sigma (C_2 -
B_2)^2$, where $C$ refers to photon counts in each bin and $B$ refers to
the background level at the same bin. The probability that the two GRBs
were both above their respective background levels and also
drawn from the same distribution was
 $$ P_{echo} = P(\chi^2_{12}) [1 - P(\chi^2_{1B})] [1 - P(\chi^2_{2B})] ,
 \eqno(1)$$
where $P(\chi^2_{12})$ is the probability that both GRBs were drawn from
the same parent distribution, given a reasonable range of $f$ and $\Delta t$,
$P(\chi^2_{1B})$ is the probability that burst \# 1 was drawn from the
background distribution, and $P(\chi^2_{2B})$ is the probability that burst
\# 2 was drawn from the background distribution.  The maximum $P_{echo}$
value was then recorded.

Further study includes comparing the GRB light curves in each of the 4
energy bands of BATSE's large area detectors.  As gravitational
lensing effects are independent of wavelength, each of the energy bands
must show acceptable echo fits for the same scale factor $f$ and time
offset $\Delta t$, independently.  In other words, the two GRBs must have
indistinguishable hardness ratios as a function of time. In addition, the
time streams before and after each GRB are checked for precursor or
post-event emission.  Such events must occur for both GRBs consistently for
the pair to be a lensing candidate.

\bigskip

\cl{\bf 3. RESULTS OF THE LENS SEARCH }

\bigskip

No clear gravitational lens examples were found.  More specifically, no two
GRB time profiles were deemed both bright enough for adequate statistical
comparison and identical in a statistical comparison.  A ``library of close
calls" (LOCC) was founded into which pairs of GRB light curves with
subjectively determined coincidental similarities were placed.  This
library serves the function of allowing us to estimate the background
against which a real gravitational lens event must be judged.  Most of the
LOCC members are bursts where one or both is quite dim, quite short in
duration, or quite featureless in appearance.

Only one pair from the Phase I search was deemed significantly similar to
suggest a detailed statistical comparison. The light curves of these two
GRBs, BATSE trigger numbers 788 and 1308, are shown in Figure 1a and 1b.
These GRBs are 26 degrees apart. Their combined positional errors are
such that if they were $11^o$ apart, there would be approximately 31.8 \%
chance that they could have come from the same position.  The positional
errors are not well described by a Gaussian distribution at large angles,
but we estimate that there is about a 10 \% chance that the two GRBs could
have come from the same position. These GRBs occurred roughly 131 days
apart. The $P(\chi^2_{echo})$ statistic was computed for a range of time
offsets and dynamic ranges to see if these two GRBs had light curves that
were consistent with being drawn from the same parent distribution.  Figure
1c shows that $P(\chi^2_{echo})$ values were never significant, and so
these GRBs cannot be considered a candidate pair for gravitational lensing.

The second LOCC entry we show here is from the Phase II search.  The GRB
pair shown in Figure 2 is particularly interesting as they each are highly
fluent, well resolved in time, and show similar although relatively smooth
time structure over their whole duration.  The positions are 3.7 degrees
apart on the sky, corresponding to 0.6 $\sigma$ of their combined
positional error, and the events are separated by 55 days. This pair,
however, is an example of how relatively featureless GRBs can fool our
search criteria. Here we show all four channels of each GRB as well as the
$P(\chi^2_{echo})$ for each channel in Figures 2a-2l. In several channels,
the two GRBs show similarities. But upon inspection of the time profiles in
different energy bands, it is clear that they have different spectra.  One
can see from the fits that no single $f$ factor is implied, and the widely
different $f$ factors clearly exclude a gravitational lens origin for the
similarity of these GRBs.

Last we discuss two GRBs that appear similar to the eye but not to a
statistical comparison.  The two GRBs with BATSE triggers 1085 and 1141
arrived 14 days apart.  They are separated by 12 degrees on the sky,
which again corresponds to only about a 10 \% chance that the two GRBs
could have originated from the same location due to their combined
positional errors.  Both light curves show similar gross features, but 1141
is actually both significantly longer and more complex than 1085.  A
statistical comparison of the light curves finds that $P_{echo}$ never even
reaches $10^{-10}$. These GRBs are illustrated in Figure 3 to demonstrate
how different bright GRBs are, in general, from each other, even after they
have been selected for similarity of appearance!

\bigskip

\cl{\bf 4. DISCUSSION AND CONCLUSIONS }

\bigskip

Should we have expected to find a lensed echo in the present search? The
smallest number of BATSE detected GRBs that Mao (1992) predicted would be
needed to be inspected to show a gravitational lens effect was 250.
Including all the limiting constraints listed in the last section, we
concluded even though 611 GRBs were compared, only effectively 50 GRBs
would be considered genuine candidates to show such an event. Therefore, it
is not surprising that we have not found a lens effect yet, and the lack of
an echo cannot be used as evidence against a cosmological setting for GRBs.
As pointed out in Wickramasinghe et al. (1993), the probability of lensing
is a very strong function of the redshift distribution of GRBs, which is
probably only weakly determined by modeling the brightness distribution.
Measurements consistent with cosmological time dilation in GRBs (Norris et
al. 1994) appear to give a GRB redshift distribution that is consistent
with the probability bounds estimated by Mao (1992) and the null detection
found so far in this search.

There is some question as to whether a GRB lens image pair would be
believed if it were found.  As a GRB is a ``once and done" occurrence, it
is not possible to re-observe the GRB pair to test for other potentially
similar characteristics.  Certainly the impossibility of distinguishing two
very weak GRBs should not lead to the conclusion that they are artifacts of
gravitational lensing.  Also, even for brighter bursts, some profiles that
are common to many GRBs might be mistaken for a lens artifact.

Clearly, if all GRBs had intrinsically identical time profiles and
power-law spectra, it would impossible to judge lens-induced similarity by
these two attributes.  In fact, it is a statement of how different GRBs are
from each other that this search can be made at all. Toward this goal, we
feel that the present work has value in helping to understand the
background of similarities between GRBs against which a lensed pair echo
must be judged.

Indeed, so far, we have found that GRBs are like snowflakes: no two
gamma-ray bursts are alike.  We note that qualitatively, the GRB comparison
search is not much different than a QSO comparison search. QSO lens images
must have identical redshift and spectra.  Were all QSO spectra identical,
it would be more difficult to determine lens pairs. Clearly then, an
understanding of the background of similarities between QSO spectra is
important in determining which are good candidates to be lens pairs.

Is it possible that two GRBs were identical but the above described
statistical test gave an unrealistically low $P_{echo}$?  We consider it
possible but unlikely.  There are at least two reasons for a false negative
correlation.  The first is that the time offset of the GRB comparison might
have skipped over the true time offset between the GRBs.  This is
particularly problematic for GRBs described by only a few time bins, or
GRBs that vary rapidly compared to the bin size (Wambsganss 1993). Clearly,
the GRBs must be compared at a time-scale large compared to the bin size
but small compared to their variability to alleviate this problem.

Secondly, comparisons might be mistakenly pessimistic if the background
levels were significantly different during the detection times of the two
GRBs. This could be caused by different solar activity levels during GRBs,
a bright gamma ray source entering or leaving earth's shadow, or {\it
Compton} entering or leaving the tail end of the South Atlantic Anomaly
during one observation.  These effects are most apparent at lower energies,
typically visible primarily in channel 1 (25 - 50 keV), and are usually not
dominate in channel 3 (100-330 keV) where the initial comparison of GRB
light curves were made.

Lastly, local effects pertinent to the spacecraft and its environment might
create false differences between the GRBs.  More specifically, electron
precipitation events (Horack et al. 1992) could add false peaks to a light
curve, or data gaps not properly accounted for could delete peaks from a
GRBs light curve.  Such electron peaks are relatively rare, usually last
for several seconds, and can usually be seen in all detectors
simultaneously, whereas GRBs would only be seen in at most 4 detectors.
Unusual data gaps, which are also relatively rare, have been made quite
clear in the data when they occur and their effects have been removed.

Could gravitational microlensing change one light curve so that it no
longer appears similar to its gravitational lens twin?  We consider this
unlikely.  It is the smooth mass distribution in galaxies that is assumed
to give rise to the multiple images of GRBs. It is these ``macroimages"
that have been searched for here. However, galaxies have at least a
reasonable fraction of their mass in stars, and these stars may themselves
become important if at least one of them approaches the light path of a
macroimage.  This is called microlensing (Chang and Refsdal 1979, Paczynski
1986a). Most probably the macroimages involved will undergo a microlensing
effect, but the typical time-scale of such an effect is likely only on the
order of milliseconds, for typical values of galactic shear and
microlensing optical depth. Since the typical light curve we inspected was
binned with 64 millisecond bins, a microlensing effect would not be
evident.  Only in the unlikely event that the optical depth is very near
unity would stellar induced microimages widely separated across the lensing
galaxy have power (Paczynski 1986a), possibly causing a microlensing
time-scale large enough to relatively distort the measured GRB light
curves.

Even if it is impossible to discern absolutely whether two GRBs are lensed
images of the same GRB, it might be possible statistically. Future
automated searches should not focus on only GRBs that are nearby each
other, but also correlate GRBs from all over the sky.  GRBs that are
similar but clearly originating from widely separated positions on the sky
-- such that lensing can not be attributed to a galaxy -- should be more
likely than those occurring near each other, and this can be tested
statistically.

It is possible, even likely, that were one GRB image seen, one or all of
its gravitational lens counterparts would be completely missed by BATSE.
Conservatively, 40 \% would be hidden from BATSE by the Earth when the
second image came in, and another 30 \% would be lost due to BATSE not
being ready to record this GRB information when it came in, due to various
gaps in experiment or satellite coverage. The combined error boxes were
searched to angular separations such that, including both random and
systematic errors, only 45 \% of lens pairs would have been found.  BATSE
is not uniformly sensitive to GRBs from all directions. Therefore there
could be GRB echoes that BATSE would not detect because the satellite was
oriented non-optimally.  We estimate that about 30 \% of GRB echoes would
be lost to this effect. We estimate that an additional 10 \% of GRBs are
too dim and of such simple structure that a statistical comparison with
other GRBs would not be definitive.  Some fraction of GRB counter-images
would be missed because either the first image arrived before BATSE
originally turned on, or the later image has yet to arrive. This last
fraction is a function of lens parameters.

Although no specific cases were discussed here, several cases have been
occasionally discussed where two images appear similar in shape but have
markedly different time-scales between them (Desai 1993).  Generally, it is
impossible to create two images with markedly different time-scales through
gravitational lensing.  The only such scenario we can envision would
require source emission angle to be strongly correlated with relativistic
beam velocity. Then one lensed image might be seen at one relativistic
$\gamma$ factor while other images might be caught at much different
$\gamma$ factors, creating the discussed effect.  Although searching for
relatively stretched images is outside the capabilities for this reported
search, no concrete case of such an effect has yet been called to our
attention that has passed our criteria for a gravitational lens image pair,
even when allowing for an arbitrary ``stretching" factor.

An automated search procedure is being designed and should begin being
implemented by the time this paper goes to press. We are hopeful this
procedure will find a gravitational lens echo, although again it is still
somewhat unlikely.

\bigskip

We thank Jerry Bonnell and Geoff Pendleton for discussions and comments,
and an anonymous referee for helpful comments. This work was supported in
part by NASA under the {\it Compton} Gamma-Ray Observatory Guest
Investigator Program.

\vfill\eject

\cl{\bf REFERENCES }
{
\parindent=0pt
\hangindent=20pt
\baselineskip=10pt
\parskip=4pt

\bigskip

Blaes, O. M. \& Webster, R. L 1992, ApJ, 391, L63
%Using gamma-ray bursts to detect a cosmological density of compact objects

\hangindent=20pt
Blandford, R. and Narayan, R., 1992, Ann. Rev. Astron. Astrophys. 30, 311

\hangindent=20pt
Brock, M. N., Meegan, C. A., Fishman, G. J., Wilson, R. B., Paciesas, W.
S., and Pendleton, G. N. 1992, in Gamma-Ray Bursts, AIP Conference
Proceedings 265, eds. W. S. Paciesas \& G. J. Fishman, (AIP, New York), 399

Chang, K. and Refsdal, S. 1979, Nature, 282, 561

Desai, U. D. 1993, private communication

Grossman, S. A., and Nowak, M. A. 1994, ApJ, submitted

\hangindent=20pt
Horack, J. M., Fishman, G. J., Meegan, C. A., Wilson, R. B., and Paciesas,
W. S. 1992, in Gamma-Ray Bursts, AIP Conference Proceedings 265, eds. W. S.
Paciesas \& G. J. Fishman, (AIP, New York), 373

Mao, S. 1992, ApJ, 389, L41
%Gravitational lensing, time delay, and gamma-ray bursts

\hangindent=20pt
Meegan, C. A., Fishman, G. J., Wilson, R. B., Paciesas, W. S., Brock, M.
N., Horack, J. M., Pendleton, G. N., \& Kouveliotou, C. 1992, Nature, 355,
143

Narasimha, D. 1993, private communication

Narayan, R. \& Wallington, S. 1992, ApJ, 399, 368
%Determination of lens parameters from gravitationally lensed
%gamma-ray bursts

Nemiroff, R. J. 1993, Comments Astrophys., in press

\hangindent=20pt
Nemiroff, R. J., Horack, J. M., Norris, J. P., Wickramasinghe, W. A. D. T.,
Kouveliotou, C., Fishman, G. J., Meegan, C. A., Wilson, R. B., and
Paciesas, W. S. 1993a, Compton Gamma-Ray Observatory, St. Louis, AIP
Conference Proceedings 280, eds. M. Friedlander, N. Gehrels, and D. J.
Macomb, (AIP, New York), 974

\hangindent=20pt
Nemiroff, R. J., Norris, J. P., Wickramasinghe, W. A. D. T., Horack, J. M.,
Kouveliotou, C., Fishman, G. J., Meegan, C. A., Wilson, R. B., and
Paciesas, W. S. 1993b, ApJ, 414, 36
% Searching GRBs for GL Echoes: Implications for Compact DM

Norris, J. P., Share, G. H., Messina, D. C., Cline, T. L., Desai, U. D.,
Dennis, B. R., Matz, S. M., \& Chupp, E. L. 1985, ApJ 301, 213

\hangindent=20pt
Norris, J. P., Nemiroff, R. J., Scargle, J. D., Kouveliotou, C., Fishman,
G. J., Meegan, C. A., Paciesas, W. S., and Bonnell, J. T. 1994, ApJ, in
press

Nowak, M. A. and Grossman, S. A. 1994, preprint

Paczynski, B. 1986a, ApJ, 301, 503

Paczynski, B. 1986b, ApJ, 308, L42
%Gamma-ray bursters at cosmological distances

Paczynski, B. 1987, ApJ, 317, L51
%Gravitational microlensing and gamma-ray bursts

Wambsganss, J. 1993, ApJ, 406, 29

\hangindent=20pt
Wickramasinghe, W. A. D. T., Nemiroff, R. J., Norris, J. P., Kouveliotou,
C., Fishman, G. J., Meegan C. A., Wilson, R. B., and Paciesas, W. S. 1993,
ApJ, 411, L55

% The Consistency of Standard Cosmology and the BATSE Number Versus
% Brightness Relation

}

\vfill\eject

\cl{\bf FIGURE CAPTIONS }
\hangindent=0pt
\baselineskip=18pt

\bigskip

\noindent {\bf Figure 1:} Light curves and probability of echo results for
two BATSE GRBs with trigger numbers 788 and 1308.  These GRBs
are not statistically identical enough to warrant a gravitational lens
interpretation for their similarity.

\bigskip

\noindent {\bf Figure 2:} Light curves and probability of echo results for
two BATSE GRBs with trigger numbers 1733 and 1956.  These GRBs are compared
in all four energy channels.  Although these high fleunce GRBs have marked
similarities, their significantly different hardness ratios exclude a
gravitational lens interpretation.

\bigskip

\noindent {\bf Figure 3:} Light curves and probability of echo results for
two BATSE GRBs with trigger numbers 1085 and 1141.  These relatively bright
GRBs have similar shapes but are not statistically comparable.  Most bright
GRBs are not even this similar, however.

\vfill\eject

\end